# Multi-User SISO Precoding based on Generalized Multi-Unitary Decomposition for Single-carrier Transmission in Frequency Selective Channel

Wee Seng Chua[1], Chau Yuen[2,3], Yong Liang Guan[1] and Francois Chin[2]
[1]Nanyang Technological University, [2]Institute for Infocomm Research, Singapore, [3]Hong Kong Polytechnic University
Email: chua0159@ntu.edu.sg, cyuen@i2r.a-star.edu.sg, eylguan@ntu.edu.sg, and francoischin@i2r.a-star.edu.sg

*Abstract* – In this paper, we propose to exploit the richly scattered multi-path nature of a frequency selective channel to provide additional degrees of freedom for desigining effective precoding schemes for multi-user communications. We design the precoding matrix for multi-user communications based on the Generalized Multi-Unitary Decomposition (GMUD), where the channel matrix H is transformed into $\mathbf{P}_i \mathbf{R}_r \mathbf{Q}_i^H$. An advantage of GMUD is that multiple pairs of unitary matrices $\mathbf{P}_i$ and $\mathbf{Q}_i$ can be obtained with one single $\mathbf{R}_r$. Since the column of $\mathbf{Q}_i$ can be used as the transmission beam of a particular user, multiple solutions of $\mathbf{Q}_i$ provide a large selection of transmission beams, which can be exploited to achieve high degrees of orthogonality between the multipaths, as well as between the interfering users. Hence the proposed precoding technique based on GMUD achieves better performance than precoding based on singular value decomposition.

## I. INTRODUCTION

MIMO technology has been widely researched in recent years due to its advantages of increasing capacity and diversity gain. With the availability of channel state information (CSI) at the transmitter, the multi-antenna techniques can be used to serve multiple users simultaneously, this is because multiple antennas at the transmitter provide additional degrees of freedom in the spatial domain. It has been shown in [1, 2] that the throughput can be increased linearly with minimum number of transmit antennas and number of users. Since then, many pre-equalization or precoding techniques have been investigated to enable multi-user MIMO broadcast channel.

Dirty paper coding [3] can be applied at the transmitter to suppress interference in multi-user MIMO systems [4]. Due to the complexity of dirty paper coding, some simple precoding techniques such as zero-forcing and regularized-inverse precoding have been developed in [5]. These precoders work well when combined with vector perturbation [6-8].

Unfortunately, most of the precoding techniques have been designed for flat fading channel, whereas a practical broadband channel almost always involves frequency selective fading. Instead of using multi-carrier signalling to convert a frequency selective fading channel into flat fading channels, we propose an alternative approach to exploit the multi-path nature of the frequency selective channel, and use the extra degrees of freedom provided by the multi-path as multiple transmit antennas to provide simultaneous transmission to multiple users.

That is, in this paper, we apply the MIMO precoding concept to service multiple users with single-carrier transmission in frequency selective channel, assuming that all the users are equipped with only single-antenna transmitter and receiver (i.e. each user channel are SISO). The objective is to use multi-path to provide the same multiplexing gain to serve multiple user in the same way as multiple antennas do. In multi-user precoding communication, interference between the multipaths can be enormously huge due to non-orthogonality of the transmission beams and multipath. In order to mitigate such interference more effectively, we propose to use a decomposition method called Generalized Multi-Unitary Decomposition (GMUD) that has been previously proposed for multi-antenna system [9] to pre-equalize the channel and at the same time, maximize orthogonality among users.

## II. SIGNAL MODEL

The system considered has one transmit antenna at the base station servicing one data stream to each of a pool of $K$ users with one receive antennas each. The received signal vector at $k^{th}$ user whose element represents the receive signal at different time interval is given as

$$\mathbf{y}_k = \mathbf{H}_k \mathbf{x} + \mathbf{n}_k \quad (1)$$

where **x** represents the complex envelopes of the transmitted signal vector, $\mathbf{H}_k$ is considered as a linear time-invariant $M$ paths fading channel, and $\mathbf{n}_k$ is the additive white Gaussian noise vector with variance $\sigma^2$. For a linear time-varying channel with a finite memory/path, we assume $M \geq K$, and we precode every $PT$ time period, where $P = (2M-1)$ and $T$ refers to the signal period. The channel matrix **H** considered in (1) is a $(2M-1) \times M$ block Toeplitz matrix:

$$\mathbf{H}_k = \begin{pmatrix} h_k[1] & 0 & \cdots & 0 \\ \vdots & \vdots & \ddots & \vdots \\ h_k[M] & h_k[M-1] & \cdots & h_k[1] \\ \vdots & \vdots & \ddots & \vdots \\ 0 & 0 & \cdots & h_k[M] \end{pmatrix} \quad (2)$$

where $h_k[m]$ represents the *m*-th path channel impulse response between the transmit and receive antennas of user *k*. The transmitter signal vector **x** can be represented as

$$\mathbf{x} = (\mathbf{Gu})/\sqrt{\gamma} \quad (3)$$

where **G** is a $M \times K$ precoding matrix, $\mathbf{u} = [u_1 \cdots u_K]^T$ whose element $u_k$ is the intended data vector for $k^{th}$ user, and $\gamma = \|\mathbf{Gu}\|^2$ is used to normalize the transmitted signal. For this paper, we limit the investigation by using linear minimum mean square error (MMSE) receiver as it is an optimum receiver for a MIMO system as shown in [10, 11].

The equivalent channel at receiver is

$$\tilde{\mathbf{H}}_k = \mathbf{H}_k \mathbf{G}/\sqrt{\gamma} \quad (4)$$

and the estimated signal for $u_k$ with MMSE receiver is

$$\hat{u}_k = \left(\tilde{\mathbf{H}}_k^H \tilde{\mathbf{H}}_k + \sigma^2 \mathbf{I}\right)^{-1} \tilde{\mathbf{H}}_k^H u_k + \left(\tilde{\mathbf{H}}_k^H \tilde{\mathbf{H}}_k + \sigma^2 \mathbf{I}\right)^{-1} \mathbf{n}_k \quad (5)$$

## III. GENERALIZED MULTI-UNITARY DECOMPOSITION (GMUD)

Given a complex $P \times M$ matrix channel **H**, where $K \leq \min(P,M)$, it can transform into various forms using different decomposition techniques such as Singular Value Decomposition (SVD) [12], Geometric Mean Decomposition (GMD) [13,14], Geometric Triangular Decomposition (GTD) [15], and Generalized Multi-Unitary Decomposition (GMUD) [9]. All the decomposition techniques mentioned above transform **H** to $\mathbf{URV}^H$, where **U** and **V** are unitary matrices, and matrix **R** varies with different decomposition.

For Generalized Multi-Unitary Decomposition (GMUD) in [9], it transforms **H** to $\mathbf{P}_i \mathbf{R}_r \mathbf{Q}_i^H$, where $\mathbf{R}_r$ is $P \times M$ matrix containing a block $K \times K$ lower triangular matrix or a special $K \times K$ matrix with a prescribed value at the first element and zeros for the rest of the elements in the first row, $\mathbf{P}_i$ and $\mathbf{Q}_i$ is a pair from a large pool of different pairs of unitary matrices. It is a general decomposition method that includes SVD, GMD, and GTD as part of the solutions. Consider a complex matrix $\mathbf{H} = \mathbb{C}^{P \times M}$, where $K = M < P$ with singular values $\lambda_1 \geq \cdots \geq \lambda_K$. $\mathbf{R}_r$ can be defined as the following form:

$$\mathbf{R}_r = \begin{bmatrix} \tilde{\mathbf{R}} \\ 0 \end{bmatrix}_{P \times M} \quad (6)$$

where

$$\tilde{\mathbf{R}} = \begin{bmatrix} r & 0 & \cdots & 0 \\ z_{21} & z_{22} & \cdots & z_{2R} \\ \vdots & \vdots & \vdots & \vdots \\ z_{R1} & z_{R2} & \cdots & z_{RR} \end{bmatrix}_{M \times M} \text{ or } \begin{bmatrix} r_1 & 0 & \cdots & 0 \\ z_{21} & r_2 & 0 & 0 \\ \vdots & \vdots & \vdots & \vdots \\ z_{R1} & z_{R2} & \cdots & r_R \end{bmatrix}_{M \times M}$$

For the first form, the non-zero positive element *r* at the (1,1) position can be assigned to any value between the largest and smallest singular value of **H**. The remaining elements at the other rows are calculated based on *r* and the singular values. For the second form, **R** is a lower triangular matrix with user pre-defined diagonal elements. This can be achieved by assigning the next remaining row in the same way after the previous row has been assigned and let all the entries on the right of the diagonal entries to be zero.

For simplicity, we consider $P = 3$ and $M = K = 2$ from this point onwards, however GMUD can be extended to different dimensions. The channel **H** is first transforms into $\mathbf{H} = \mathbf{U}\mathbf{\Lambda}\mathbf{V}^H$ using SVD. The matrix $\mathbf{R}_r$ with a pre-assigned value $\lambda_1 < r \leq \lambda_K$ at the (1,1) position can be arranged in the form of $\mathbf{R}_r = \mathbf{W}^H \mathbf{\Lambda} \mathbf{X}^H$, where **W** and **X** are unitary matrices assigned using Givens rotations, **Λ** is the same diagonal matrix of **H** containing the singular values.

$$\mathbf{W} = \begin{bmatrix} a^H & b^H & 0 \\ -b^H & a^H & 0 \\ 0 & 0 & 1 \end{bmatrix}, \mathbf{X} = \begin{bmatrix} c & s \\ -s & c \end{bmatrix} \quad (7)$$

As a result, *r*, $z_1$ and $z_2$ can be defined in terms of *a, b, c, s,* $\lambda_1$ and $\lambda_2$.

$$\mathbf{\Lambda} = \mathbf{W} \mathbf{R}_r \mathbf{X}^H$$
$$\mathbf{R}_r = \mathbf{W}^H \mathbf{\Lambda} \mathbf{X}$$

$$\begin{bmatrix} r & 0 \\ z_1 & z_2 \\ 0 & 0 \end{bmatrix} = \begin{bmatrix} a^H & b^H & 0 \\ -b^H & a^H & 0 \\ 0 & 0 & 1 \end{bmatrix}^H \begin{bmatrix} \lambda_1 & 0 \\ 0 & \lambda_2 \\ 0 & 0 \end{bmatrix} \begin{bmatrix} c & s \\ -s & c \end{bmatrix} \quad (8)$$

$$\begin{bmatrix} r & 0 \\ z_1 & z_2 \\ 0 & 0 \end{bmatrix} = \begin{bmatrix} ac\lambda_1 + bs\lambda_2 & as\lambda_1 - bc\lambda_2 \\ bc\lambda_1 - as\lambda_2 & bs\lambda_1 + ac\lambda_2 \\ 0 & 0 \end{bmatrix}$$

The value of *a* and *c* can be derived from the first row of (8), after substituting *b* and *s* from (7), and subsequently **W** and **X** can be determined.

$$ac\lambda_1 + bs\lambda_2 = r, \quad as\lambda_1 - bc\lambda_2 = 0 \quad (9)$$

The remaining values of $\mathbf{R}_r$ can be found by substituting the values of *a* and *c*

$$z_1 = bc\lambda_1 - as\lambda_2, \quad z_2 = bs\lambda_1 + ac\lambda_2 \quad (10)$$

Subsequently, **H** can be decomposed into

$$\mathbf{H} = \mathbf{U}\mathbf{W}\mathbf{R}_r \mathbf{X}^H \mathbf{V}^H = (\mathbf{U}\mathbf{W})\mathbf{R}_r (\mathbf{V}\mathbf{X})^H = \mathbf{P}_r \mathbf{R}_r \mathbf{Q}_r^H \quad (11)$$

$\mathbf{P}_r$ and $\mathbf{Q}_r$ are unitary matrices due to the combination of multiple unitary matrices.

In order to get multiple different **P** and **Q**, we include phase rotation matrices $\mathbf{M}_{\theta_1}$ and $\mathbf{M}_{\theta_2}$ to (11). $\mathbf{M}_{\theta_1}$ and $\mathbf{M}_{\theta_2}$ are diagonal matrices whose elements are direction parameter $\theta$ with unity gain

$$\mathbf{M}_{\theta_1} = \begin{bmatrix} e^{j\theta} & 0 & 0 \\ 0 & 1 & 0 \\ 0 & 0 & 1 \end{bmatrix} \text{ and } \mathbf{M}_{\theta_2} = \begin{bmatrix} e^{j\theta} & 0 \\ 0 & 1 \end{bmatrix} \quad (12)$$

where $\theta$ can be any value from 0 to $2\pi$. After the inclusion of $\mathbf{M}_{\theta_1}$ and $\mathbf{M}_{\theta_2}$, **H** becomes

$$\begin{aligned}\mathbf{H} &= \mathbf{U}\mathbf{M}_{\theta_1}\mathbf{\Lambda}\mathbf{M}_{\theta_2}^H \mathbf{V}^H \\ &= \mathbf{U}\mathbf{M}_{\theta_1}\left(\mathbf{W}\mathbf{R}_r\mathbf{X}^H\right)\mathbf{M}_{\theta_2}^H\mathbf{V}^H \\ &= \left(\mathbf{U}\mathbf{M}_{\theta_1}\mathbf{W}\right)\mathbf{R}_r\left(\mathbf{V}\mathbf{M}_{\theta_2}\mathbf{X}\right)^H \\ &= \mathbf{P}_{\theta,r}\mathbf{R}_r\mathbf{Q}_{\theta,r}^H \end{aligned} \quad (13)$$

It is shown in (13), $\mathbf{R}_r$ is independent to the value of $\mathbf{M}_{\theta_1}$ and $\mathbf{M}_{\theta_2}$. From (13), it is clear that the values of $\mathbf{P}_{\theta,r}$ and $\mathbf{Q}_{\theta,r}$ change with the values of $\theta$ and $r$ in $\mathbf{R}_r$. Moreover, since $\mathbf{U}$, $\mathbf{W}$, $\mathbf{V}$, $\mathbf{X}$, $\mathbf{M}_{\theta_1}$ and $\mathbf{M}_{\theta_2}$ are unitary matrices, $\mathbf{P}_{\theta,r}$ and $\mathbf{Q}_{\theta,r}$ are unitary matrices.

## IV. PRECODING BASED ON GMUD

Consider the case where there is one transmit antenna at the base station sending one data stream to each of the $K$ users with one receive antennas each. The corresponding channel matrix $\mathbf{H}_k$ of each user as given in (1) can be decomposed using GMUD. Depending on the gain parameter $r$ in $\mathbf{R}_r$, multiple different pairs of $\mathbf{P}_{\theta,r}$ and $\mathbf{Q}_{\theta,r}$ matrices can be generated with the same $\mathbf{R}_r$ matrix, and each user's channel can be decomposed into

$$\mathbf{H}_k = \mathbf{P}_{\theta_k,r_k}\mathbf{R}_{r_k}\mathbf{Q}_{\theta_k,r_k}^H \quad (14)$$

where we replace $r$ in $\mathbf{R}$ from (6) and $\theta$ in $\mathbf{M}_{\theta_1}$ and $\mathbf{M}_{\theta_2}$ from (12) with optimizing parameters $r_k$ and $\theta_k$ respectively. The first column vector of $\mathbf{Q}_{r_k,\theta_k}$ is considered as an individual transmission beam for that user. Thus, by adjusting $r_k$ and $\theta_k$, multiple $\mathbf{Q}_{r_k,\theta_k}$ correspond to different transmission beams with different amplitude $r_k$ and different directions $\theta_k$ can be obtained. The transmitter at the base station will steer the beams of every users to ensure they are matched as orthogonal as possible. The orthogonality between the transmission beams is related to the multi-user interference, hence if all the users' transmission beams are orthogonal to each other, there will be zero multi-users interference.

The transmitter precodes the data vector by multiplying a linear precoding matrix **G** containing the first column vectors of $\mathbf{Q}_{r_k,\theta_k}$ of each users. From (13) and (14),

$$\mathbf{Q}_{r_k,\theta_k} = \mathbf{V}\mathbf{M}_{\theta_k}\mathbf{V}_0 = \begin{bmatrix} \underbrace{\mathbf{q}_{1,r_k,\theta_k}}_{\mathbf{g}_k} & \mathbf{q}_{2,r_k,\theta_k} \end{bmatrix} \quad (15)$$

where $\mathbf{q}_{1,r_k,\theta_k}$ and $\mathbf{q}_{2,r_k,\theta_k}$ are the first and second column vectors respectively. We let

$$\mathbf{g}_k = \mathbf{q}_{1,r_k,\theta_k}, \qquad \mathbf{g}_l = \mathbf{q}_{1,r_l,\theta_l} \quad (16)$$

and the precoding matrix becomes

$$\mathbf{G} = [\mathbf{g}_k \quad \mathbf{g}_l] \quad (17)$$

The unbalance channel gain of each user's transmission beam can be compensated by introducing different power loading factor to $\mathbf{g}_k$ and $\mathbf{g}_l$ and **G** becomes

$$\mathbf{G} = [\alpha\mathbf{g}_k \quad \beta\mathbf{g}_l] \quad (18)$$

When we combine (1), (14) and (17), the received signal for user $k$ becomes

$$\begin{aligned}\mathbf{y}_k &= \mathbf{P}_{\theta_k,r_k}\mathbf{R}_{r_k}\left(\mathbf{Q}_{\theta_k,r_k}\right)^H \frac{\mathbf{G}\mathbf{u}}{\sqrt{\gamma}} + \mathbf{n}_k \\ &= \mathbf{P}_{\theta_k,r_k}\mathbf{R}_{r_k}\begin{bmatrix} \mathbf{q}_{1,\theta_k,r_k}^H \\ \mathbf{q}_{2,\theta_k,r_k}^H \end{bmatrix}[\alpha\mathbf{g}_k \quad \beta\mathbf{g}_l]\frac{\mathbf{u}}{\sqrt{\gamma}} + \mathbf{n}_k \\ &= \frac{1}{\sqrt{\gamma}}\mathbf{P}_{\theta_k,r_k}\mathbf{R}_{r_k}\begin{bmatrix} \alpha\mathbf{q}_{1,\theta_k,r_k}^H\mathbf{g}_k u_k + \beta\mathbf{q}_{1,\theta_k,r_k}^H\mathbf{g}_l u_l \\ \alpha\mathbf{q}_{2,\theta_k,r_k}^H\mathbf{g}_k u_k + \beta\mathbf{q}_{2,\theta_k,r_k}^H\mathbf{g}_l u_l \end{bmatrix} + \mathbf{n}_k \end{aligned} \quad (19)$$

where $\gamma = \|\mathbf{G}\mathbf{u}\|^2$ is used to normalized the transmitted signal.

Using the matrix structure of $\mathbf{R}_k$ as shown in (6), and $\mathbf{q}_{1,r_k,\theta_k}^H\mathbf{g}_k = 1$ as shown in (16), (19) can be reduced to

$$\begin{aligned}\mathbf{y}_k &= \frac{1}{\sqrt{\gamma}}\mathbf{P}_{\theta_k,r_k}\begin{bmatrix} r_k\left[\alpha\mathbf{q}_{1,\theta_k,r_k}^H\mathbf{g}_k u_k + \beta\mathbf{q}_{1,\theta_k,r_k}^H\mathbf{g}_l u_l\right] \\ \varepsilon \\ 0 \end{bmatrix} + \mathbf{n}_k \\ &= \frac{1}{\sqrt{\gamma}}\mathbf{P}_{\theta_k,r_k}\begin{bmatrix} r_k\left[\alpha u_k + \beta\mathbf{q}_{1,\theta_k,r_k}^H\mathbf{g}_l u_l\right] \\ \varepsilon \\ 0 \end{bmatrix} + \mathbf{n}_k \end{aligned} \quad (20)$$

where $\varepsilon = \alpha z_{21,k}u_k + \left(\alpha z_{21,k}\mathbf{q}_{1,\theta_k,r_k}^H\mathbf{g}_l + \beta z_{22,k}\mathbf{q}_{2,\theta_k,r_k}^H\mathbf{g}_k\right)u_l$, and $z_{21,k}$ and $z_{22,k}$ refer to the elements of $\mathbf{R}_k$ in (6). The signal is $\alpha r_k u_k/\sqrt{\gamma}$ and the interference is $\beta r_k\mathbf{q}_{1,\theta_k,r_k}^H\mathbf{g}_l u_l/\sqrt{\gamma} = \beta r_k\mathbf{q}_{1,\theta_k,r_k}^H\mathbf{q}_{1,\theta_l,r_l}u_l/\sqrt{\gamma}$. To detect $u_k$, the user can either use a zero-forcing or MMSE receiver as shown in (5), where MMSE receiver or Wiener filter is shown [10, 11] as the optimum receiver for a MIMO system.

In order to reduce BER, $\mathbf{q}_{1,r_k,\theta_k}$ and $\mathbf{g}_l$ must be made as orthogonal as possible because the orthogonality between them contributes to the multi-user interference. However, a small dot product between these vectors may not lead to a good precoding matrix **G** because G may have a large normalization constant $\gamma$ which results in a weaker received signal and increases the probability of bit error at the receiver. Hence it is important to optimize a signal/interference power related cost function instead. Thus, in this paper, we present the use of minimizing the sum of per user's inverse SINR criteria. This is because optimizing this criterion has a better BER performance than other criteria such as maximizing minimum SINR and maximizing sum of per user's SINR. The cost function of finding **G** becomes

$$\mathbf{G} = \min_{r_i, \theta_i, \forall 1 \leq i \leq K} \sum_{k=1}^{K} (\text{SINR}_i)^{-1} \quad (21)$$

In the case of two users with two multi-path each, the cost function becomes

$$\mathbf{G} = \min_{r_k, r_l, \theta_k, \theta_l, \alpha_k, \alpha_l} \left( \frac{\alpha_l^2 \left| \mathbf{q}_{1,\theta_k,r_k}^H \mathbf{q}_{1,\theta_l,r_l} \right|^2}{\alpha_k^2} + \frac{\sigma^2 \gamma}{\alpha_k^2 |r_k|^2} + \frac{\alpha_k^2 \left| \mathbf{q}_{1,\theta_l,r_l}^H \mathbf{q}_{1,\theta_k,r_k} \right|^2}{\alpha_l^2} + \frac{\sigma^2 \gamma}{\alpha_l^2 |r_l|^2} \right) \quad (22)$$

## IV. SIMULATION RESULTS AND DISCUSSION

For simplicity, we consider one transmit antenna at the base station and one receive antenna for each user. Every user's channel contains two equal gain multi-path. Since GMUD provides multiple different transmission beams from the first column vector of multiple different unitary matrix $\mathbf{Q}_{r_k,\theta_k}$ of each users, the requirement of perfect channel state information (CSI) is not stringent. It is shown in [9] that precoding based on GMUD is a robust scheme which does not suffer much loss from limited or inaccurate feedback information.

For comparison purposes, we also consider the precoding based on SVD. We use the principle eigen-vector of the channel, which has the largest channel gain represented by the principle singular value. The linear precoding matrix at the base station contains all the users' principle eigenvectors, while the downlink users use MMSE receivers to receive and decode the received signal.

To have more insight into the behavior of GMUD in frequency selective versus frequency flat multi-user channels, we also compare the proposed SISO precoding scheme with the MIMO precoding scheme of [9]. In the latter, there are two transmitting antennas at the base station and two receiving antennas at the mobile terminals, the base station services one data stream to every user simultaneously in a flat fading channel [9].

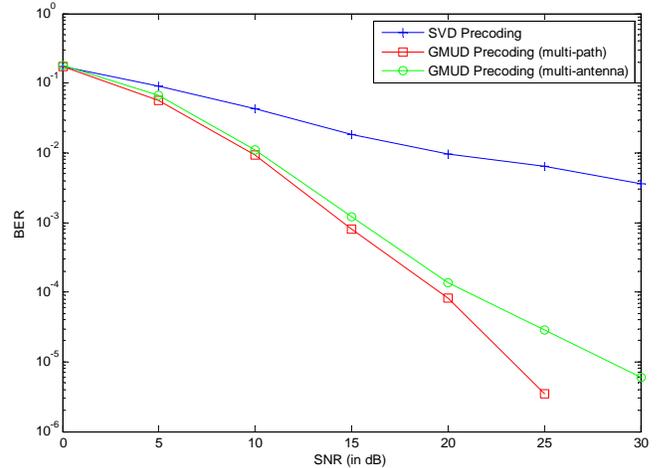

**Figure 1:** Performance of probability of bit error for two users using precoding matrix that send QPSK symbols based on GMUD and SVD respectively. The precoding matrix based on GMUD with multipath is a two path SISO frequency selective channel and GMUD with multi-antenna is a two transmit antennas and two receive antennas per user MIMO multi-user flat fading channel respectively.

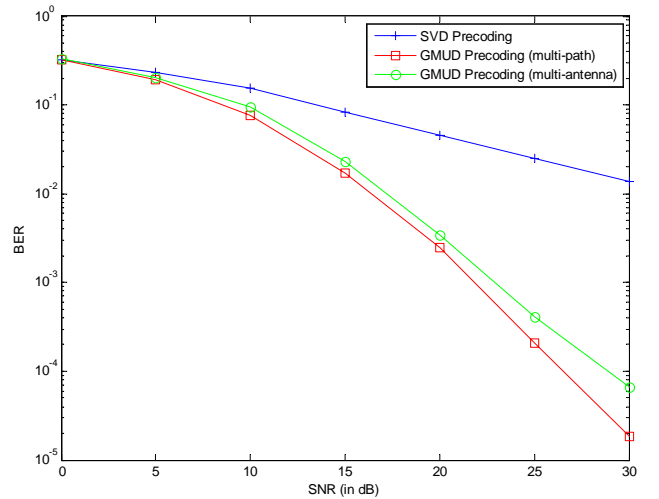

**Figure 2:** Performance of probability of bit error for two users using precoding matrix that send 16QAM symbols based on GMUD and SVD respectively. The precoding matrix based on GMUD with multiplath is a two path SISO frequency selective channel and GMUD with multi-antenna is a two transmit antenna and two receive antennas per user MIMO multi-user flat fading channel respectively.

Figure 1 and Figure 2 show that under perfect CSI at the transmitter, precoding based on GMUD has a significant gain over precoding based on SVD. This implies that although the principal SVD transmission beams of each user has the largest channel gain, collectively the interference between the users is extremely high. This is because the transmission beams for different user are not orthogonal and this give rises to crosstalk between different users. On the other hand, precoding based on GMUD chooses the transmission beams that are more orthogonal, based on optimizing the cost function in (21) or (22) from a pool of transmission beams centered at the principle eigen-vector. For more information about the direction of the first column vectors of different unitary matrix $\mathbf{Q}_{r_k,\theta_k}$, please refer to [9].

Under this condition, precoding based on GMUD ensures that both multipath and multi-user inference and noise are taken care of simultaneously.

Next, we compare the performance of GMUD precoding between a SISO frequency selective channel and MIMO flat fading channel. The degrees of freedom is similar for the two scenarios where the former scheme uses one transmit antenna at the base station and one receive antenna for every channel communicating in a two path frequency selective channel, whereas the latter scheme uses two transmit antennas at the base station and two receive antennas for every user in a flat fading channel. Figure 1 and Figure 2 show that the SISO multiple path channel results in a better performance than the MIMO flat fading channel, this is because the channel matrix with a Toeplitz form in (2) has a lower condition number. In other words, the dynamic range of $r$ in the $\mathbf{R}$ matrix of (6) becomes larger, this leads to a larger pool of $\mathbf{Q}$ matrices generated. With a larger pool of transmission beams, the optimal transmission beams used in the precoding matrix can perform better.

## V. CONCLUSION

In this paper, we propose to make use of the degrees of freedom provided by the multi-path of a frequency selective channel to support multi-user communications by precoding. Based on the precoding technique Generalized Multi-unitary Decomposition (GMUD) previously designed for flat fading MIMO multi-user system, we have applied it to frequency selective SISO multi-user system to combat the impairments induced by frequency selective fading and multiple user interference. We show that the degree of freedom offered by the multi-path can remove the need for multiple antennas for a wireless communication link that employs precoding.

## VI. REFERENCES

[1] MG. Caire and S. Shamai, "On the achievable throughput of a multiantenna Gaussian broadcast channel," *IEEE Trans. on Info. Theory*, vol. 49, no. 7, pp. 1691 – 1706 , July 2003

[2] S. Vishwanath, N. Jindal, and A. Goldsmith, "Duality, achievable rates and sum-rate capacity of Gaussian MIMO broadcast channel," *IEEE Trans. on Info. Theory*, vol. 49, no. 10, pp. 2658 – 2668, Oct. 2003

[3] M. Costa, "Writing on dirty paper," *IEEE Trans. on Info. Theory,* vol. 29, no. 3, pp. 439 – 441, May 1983

[4] M. Sharif and B. Hassibi, "A comparison of time-sharing, DPC, and beamforming for MIMO broadcast channels with many users," *IEEE Trans. on Commun.,* vol. 55, no. 1, Jan. 2007

[5] C. Peel, B. M. Hochwald, and A. L. Swindlehust, "A vector perturbation technique for near-capacity multiantenna multiuser communication – Part I: Channel inversion and regularization," *IEEE Trans. on Commun.*, vol. 51, no. 1, pp. 195 – 202, Jan. 2005

[6] B. M. Hochwald, C. Peel, and A. L. Swindlehust, "vector perturbation technique for near-capacity multiantenna multiuser communication – Part II: Perturbation," *IEEE Trans. on Commun.,* vol. 51, no. 3, pp.537 – 544, Mar. 2005

[7] W. S. Chua, C. Yuen, and F. Chin, "A continuous vector-perturbation for multi-antenna multi-user communication," *IEEE VTC Spring 2007*

[8] C. Yuen and B. M. Hochwald, "How to gain 1.5dB in vector perturbation," *IEEE Global Telecommunications Conference (Globecom)*, 2006

[9] W. S. Chua, C. Yuen, Y. L. Guan, and F. Chin, "Limited feedback for multi-antenna ulti-user communications with generalized multi-unitary decomposition," *IEEE PIMRC 2008*

[10] D. P. Palomar, J. M. Cioffi, and M. A. Lagunas, "Joint Tx-Rx beamforming design for multicarrier MIMO channels: A unified framework for convex optimization," IEEE Trans. on Signal Processing, vol. 51, no.9, pp. 2381 – 2401, Sep. 2003

[11] A. Scaglione, P. Stoica, S. Barbarossa, and H. Sampath, "Linear precoders and decoders designs for MIMO frequency selective channels," *IEEE ICASSP 2002*

[12] G. H. Golub and C. F. van Loan, Matrix Computations, 3$^{rd}$ Edition, Johns Hopkins University Press

[13] J-K. Zhang, A Kavcic, and K. M. Wong, "Equal-diagonal QR Decomposition and its Application to precoder design for successive-cancellation detection," IEEE Trans. on Info. Theory, vol. 51, pp. 154-172, Jan 2005

[14] Y. Jiang, W. W. Hager, and J. Li, "The geometric mean decomposition," Linear Algebra and Its Applications, vol. 396, pp. 373-384, Feb 2005

[15] Y. Jiang, W. W. Hager, and J. Li, "The generalized triangular decomposition," Mathematics of Computation, Nov 2006